\documentclass[twocolumn,prd,nofootinbib,aps,prl,floats,floatfix,amsmath,amssymb,longbibliography,secnumarab
ic]{revtex4-1} %
\usepackage[final]{graphicx}
\usepackage{hyperref}
\usepackage{amsmath}
\usepackage{bbm}
\usepackage{amsfonts}
\usepackage{amssymb}
\usepackage{latexsym}
\usepackage{graphicx}
\usepackage[english]{babel}
\usepackage{multirow}
\usepackage{float}
\usepackage{url}
\usepackage{slashed}
\usepackage{xcolor} 
\usepackage[utf8]{inputenc}
\usepackage{verbatim}
\usepackage{stackengine}

\def\sfrac#1#2{{\textstyle{#1\over #2}}}
\newcommand{\be}{\begin{equation}}
\newcommand{\ee}{\end{equation}}
\newcommand{\ba}{\begin{array}}
\newcommand{\ea}{\end{array}}
\newcommand{\bea}{\begin{eqnarray}}
\newcommand{\eea}{\end{eqnarray}}
\newcommand{\sss}{\scriptscriptstyle}

\newcommand{\nn}{\nonumber}

\DeclareRobustCommand{\rchi}{{\mathpalette\irchi\relax}}
\newcommand{\irchi}[2]{\raisebox{\depth}{$#1\chi$}}

\begin{document}

\title{Electroweak baryogenesis from light fermion sources: a critical study}
\author{James M.\ Cline}
\author{Benoit Laurent}
\affiliation{McGill University, Department of Physics, 3600 University St.,
Montr\'eal, QC H3A2T8 Canada}

\begin{abstract}
Electroweak baryogenesis (EWBG) is sourced by nonstandard $CP$-violating interactions of
the Higgs boson with fermions, usually taken to be the top quark, enhanced by its large Yukawa coupling.  Numerous papers have studied EWBG sourced by lighter fermions, including the tau lepton and off-diagonal quark mass terms.  We critically reassess the viability of EWBG in these scenarios, comparing the predictions based on the semiclassical (WKB) formalism for the source term to those from the VEV insertion approximation (VIA), using updated values for the collision terms, and clarifying discrepancies in the definition of the weak sphaleron rate.  The VIA systematically predicts a baryon asymmetry that is orders of magnitude larger than the WKB formalism.
We trace this to the differing shapes of the $CP$-violating source terms in the two formalisms, showing that the additional spatial derivative in the WKB source term causes large cancellations when it is integrated over the bubble wall profile.  An important exception is a source term from $c$-$t$ quark mixing, where the WKB prediction also allows for a realistically large baryon asymmetry.
In contrast, the analogous $b$-$s$ mixing source is found to be orders of magnitude too small.

\end{abstract}
\maketitle

\section{Introduction}

Electroweak baryogenesis (EWBG) has been extensively studied since its introduction
\cite{Bochkarev:1990fx,Cohen:1990py,Cohen:1990it}, in part because of its highly predictive nature.  Requiring new physics near the TeV scale, particle physics models incorporating EWBG typically predict new states close to the sensitivity of limits from the Large Hadron Collider (LHC) or searches for electric dipole moments.  However after three decades, there persists a large systematic uncertainty in the prediction of the baryon asymmetry, due to the existence of two competing formalisms for the source term that encodes $CP$-violating interactions in the bubble walls during the first-order electroweak phase transition.  The two methods are known respectively as the
VEV-insertion approximation (VIA) \cite{Riotto:1995hh,Riotto:1997vy} and the semiclassical or WKB approximation \cite{Joyce:1994fu,Joyce:1994zt,Cline:2000kb}.  Both frameworks agree that a spatially-varying
$CP$-violating phase should be present within the bubble wall in order to generate a nonvanishing baryon asymmetry of the universe (BAU), but the WKB source term involves an additional spatial derivative relative to the VIA source, leading to significant quantitative differences in the predicted BAU.

These differences were noticed in studies of EWBG in the minimal supersymmetric
standard model \cite{Carena:1997gx,Cline:1997vk,Cline:2000nw}, 
but no systematic comparison between the two methods, as a function of bubble wall parameters, was made until Ref.\ \cite{Cline:2020jre}, in the context of an extra singlet scalar coupling to the Higgs and top quark fields.  There it was found that the predictions typically differed by factors of 10-40, depending on the bubble wall speed and thickness.

The two formalisms also agree that the source term is proportional to the
mass squared of the particle that is undergoing $CP$-violating interactions in the wall.  On this basis, one would expect that the BAU induced by light fermion sources would pay the penalty of the small mass.  However, not only is the size of the source term important, but also how far the particle is able to diffuse in the plasma in front of the bubble wall, where baryon-violating sphaleron interactions are fast.  Hence EWBG induced by $CP$-violating
$\tau$ lepton interactions has also been considered, since leptons have a much longer diffusion length than quarks \cite{Joyce:1994bi,deVries:2018tgs,Xie:2020wzn}.  It has been found in the VIA approach that this can give successful EWBG, but the scenario has never been considered within the semiclassical treatment.\footnote{The seminal ref.\  \cite{Joyce:1994bi} used only a phenomenologically motivated source term, quite different from the rigorously derived one.}  Filling this gap is one aim of the present work.

In a similar vein, off-diagonal quark mass matrix elements can provide a
$CP$-violating source for EWBG in two Higgs doublet models (2HDMs).  This has been considered for $b$-$s$ \cite{Liu:2011jh,Modak:2018csw} and $t$-$c$ mixing \cite{Fuyuto:2017ewj}.  Like the tau lepton source, these scenarios have not been previously considered within the WKB approach.  We will show that it predicts too small a BAU compared to the observed value in the $b$-$s$ mixing model, but fortuitously the WKB formalism agrees that a large enough BAU is possible in the $t$-$c$ mixing scenario.

The disparate formalisms for the source terms are historically also tied to
different forms of the Boltzmann equations that propagate the source into particle asymmetries in the vicinity of the bubble wall. Namely, the VIA is always formulated together with single-component diffusion equations, whereas
WKB is implemented with two moments of the Boltzmann equation for each particle species.  However there is no absolute need for these associations, and part of our comparison will be to study the effect of the approximation made to the full Boltzmann equations, independently of the choice of source term.

We begin in Section \ref{sect:sources} by reviewing the two formalisms for the sources and the fluid equations, and the means for transforming the WKB or VIA source terms so that they can be used in the fluid equation network of the opposing framework.  At the same time, we will review and update the collision rates for the various relevant processes, to advocate a preferred set of values that should be used in either formalism.  Differing choices are a further source of discrepancies in the literature that we strive to eliminate in this work.

In Section \ref{sect:discrepancy} we carry out an analytic approximation for the BAU predicted by the two formalisms, using
the Green's function approach, to give insight into the parametric differences between the predictions.  In the remainder of the paper, fully numerical solutions of the fluid equations are implemented to validate these results, in the case of a tau
lepton source (Section \ref{sect:tau}) and $c$-$t$ or $b$-$s$ quark mixing sources (Section \ref{sect:mixing}).
A summary and conclusions are given in Section \ref{sect:concl}.

\begin{figure}[!tb]
\begin{center}
\centerline{
\includegraphics[scale=0.35]{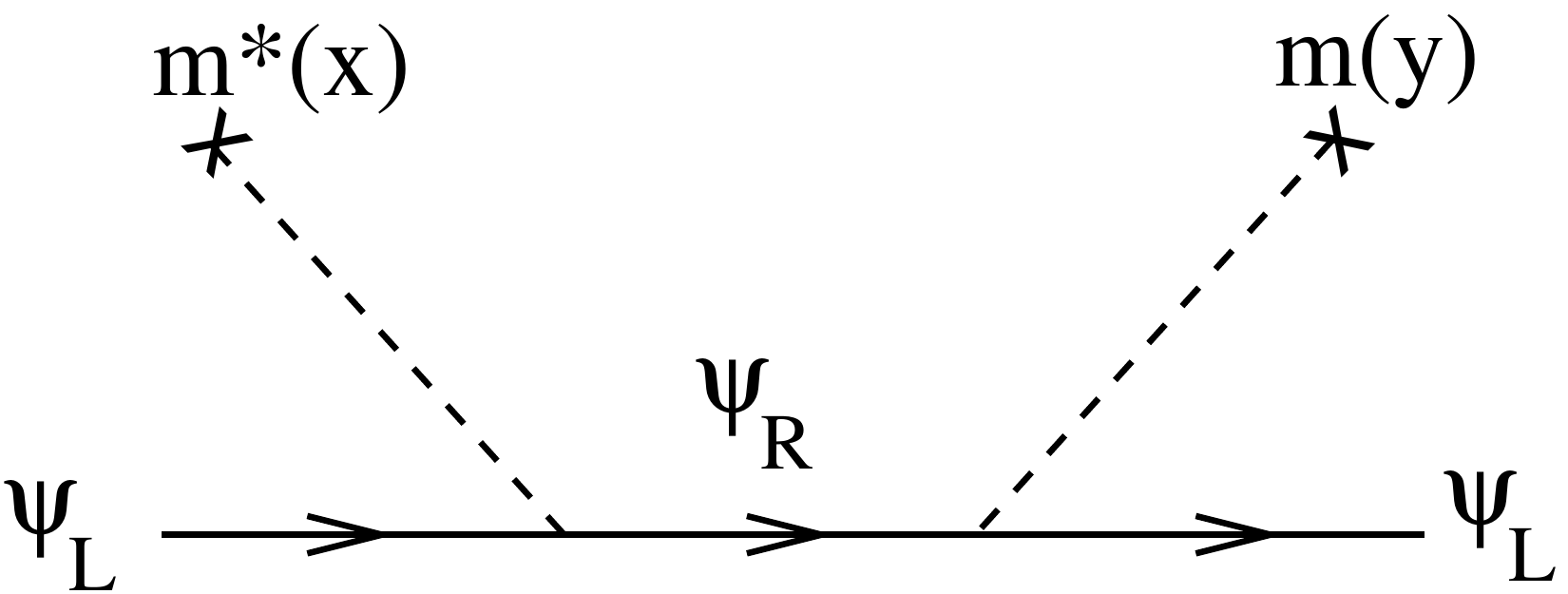}}
 \caption{Perturbative contribution to fermion self-energy
 $\Sigma(x,y)$ used in VIA source computation.} 
 \label{fig:sigma}
\end{center} 
\end{figure}

\section{Sources and fluid equations}
\label{sect:sources}
We begin by briefly reviewing the origin and derivation of the source terms and fluid equations in the two competing frameworks.  These determine the particle asymmetries that eventually bias sphalerons and lead to the baryon asymmetry.  

In the following we will assume that the generation of the initial asymmetries occurs on a small timescale compared to the weak sphaleron interactions, which allows one to split the BAU computation into two steps: first compute the chiral asymmetries, and then take these as inputs to the Boltzmann equation for baryon production.  One could alternatively include weak sphalerons in the collision terms and add an extra equation for baryon production, solving the whole system simultaneously.  It has been shown that either method gives the same results to within a few percent \cite{deVries:2018tgs}.

\subsection{VIA formalism}
The starting point for the VIA method is a set of Boltzmann equations for the particle asymmetries $n_i$, originally set out in Ref.\ \cite{Huet:1995sh}. 
The asymmetries are parametrized as $n_i = k_i \mu_i T^2/6$, where
$\mu_i$ is the chemical potential and $k_i$ counts degrees of freedom.
 The equations take the generic form
\be
    \dot n_i = D_i \partial_z^2 n_i -{\cal C}_{i}^{\sss V\!I\!A}[n_j] + S_{{\sss V\!I\!A},i}
\label{HNeq}
\ee
where $D_i$ is the diffusion constant, ${\cal C}_{i}^{\sss V\!I\!A}$ is the total collision rate, proportional to  a signed linear combination of densities corresponding to particles in the external states (including species $i$), and $S_i$ is an inhomogeneous source term.  In the rest frame of the bubble wall, and for steady-state solutions, one makes the replacement
$\dot n_i\to v_w \partial_z n_i$, where $v_w$ is the wall velocity.
Phenomenologically, the source term $S_{{\sss V\!I\!A},i}$ can be related to the nonconservation of the $n_i$ number density by considering the current $J_i^\mu$ corresponding to $n_i = J_i^0$;  namely, $\partial_\mu J_i^\mu = 
\dot n_i - \vec\nabla\cdot\vec J_i  = S_{{\sss V\!I\!A},i}$.  This matches the form of Eq.\ (\ref{HNeq}) (ignoring the collision term $\Gamma_i$) if Fick's law is approximately satisfied, $\vec J_i = D\vec\nabla n_i$, and if $S_{{\sss V\!I\!A},i} = 
\partial_\mu J_i^\mu$.

In the seminal references \cite{Riotto:1995hh,Riotto:1997vy}  for the VIA, it was proposed to compute $\partial_\mu J^\mu_i$ perturbatively in a
lowest-order expansion in the background fields of the bubble wall,
that give rise to a $CP$ violating source, using finite-temperature
field theory in the closed-time-path  (CTP) formalism due
to Schwinger and Keldysh; see {\it e.g.,} \cite{Henning:1995sm}.  For example, a fermion has $J^\mu(x) = \bar\psi(x)\gamma^\mu \psi(x)$, and
$\partial_\mu \langle J^\mu\rangle$ can be related in thermal field theory to propagators $S$ and self-energies $\Sigma$ in position space, using the Schwinger-Dyson
equation, with the form
\bea
\partial_\mu \langle J^\mu(x)\rangle &=&  \int d^{\,4}y \,{\rm tr}
\Big( S^>(x,y)\Sigma^<(y,x) +\{x\leftrightarrow y\}
\Big)\nn\\
&-& \{\Sigma \leftrightarrow S\}\,.
\label{eq:divJ}
\eea
The VIA consists in approximating the self-energy perturbatively, by 
expanding in the leading interactions that give $\Sigma$ in terms of the background fields in the wall, and neglecting the background fields in the propagators. 

An example is given in Figure \ref{fig:sigma}, where
the leading contribution to $\Sigma$ for one chirality of $\psi$ is shown schematically, assuming the $CP$-violation is due to insertions of background spatially varying mass terms $m(x) = y_1 v_1 + y_2 v_2$ in a 2HDM with complex Yukawa couplings.  In this case the source
involves 
\bea\label{eq:VIAsource}
S_{\rm\sss VIA}(x) &\sim& \int d^{\,4}y\, \left(m_xm^*_y-m^*_x m_y\right){\rm Im\, tr}\left[S^>_{xy}S^<_{yx}\right]\nn\\
    &\sim& {\rm Im}(y_1 y_2^*) v_2^2 (v_1/v_2)',
\eea
where prime denotes $d/dz$ in the bubble wall and $m_x \equiv m(x)$.
A peculiarity of the VIA is that the numerical prefactor of the source term, which involves a phase space integral over the thermal distribution function of the internal fermion, is infrared divergent unless the thermal width $\Gamma_T$ of the fermion is taken into account 
\cite{Lee:2004we}.  Therefore the VIA source diverges in the $T\to 0$ limit.  Although this is not a practical limitation, it raises concerns at the conceptual level.

{Recently, Ref.\ \cite{Postma:2021zux} argued that the VIA is only valid for $L_w\gg\Gamma_T^{-1}$, hence the aforementioned IR divergence is an artifact of going beyond the regime of validity of the derivative expansion around
$y=x$, needed to extract the source (\ref{eq:VIAsource}) from the current (\ref{eq:divJ}).  This condition is particularly restrictive for leptons, which have $\Gamma_T = 0.002T$ \cite{Elmfors:1998hh}. It imposes a lower bound on $L_w$ that far exceeds the typical thickness of $L_w\sim10/T$ \cite{Cline:2021iff} of a realistic profile,
bringing into question the validity of VIA for models
with a $CP$-violating source from leptons.}

A further concern is that since VIA is an expansion in powers of $m_i^2/T^2$, it is not clear that the leading contribution can be trusted for heavy particles.  In fact ref.\ \cite{Postma:2019scv} has shown that the next-to-leading correction dominates for top quark sources, casting doubt on the validity of VIA for heavy particle sources.

It is worth emphasizing that the VIA diffusion equations are not,
strictly speaking, derived from the CTP formalism, as is often stated.  Rather it is the divergence of a current that is derived, and this divergence equation is reinterpreted in a phenomenological way as a Boltzmann equation, by invoking Fick's law for the spatial current density.

\subsection{WKB formalism}
Historically, the starting point for deriving the WKB fluid equations
is the Boltzmann equation for the asymmetries between the particle and antiparticle distribution functions, 
$\delta f_i(x,p)$, including a force term, that can encode the force $\vec F$ of the bubble wall acting on the particles. 
Splitting $\vec F$ into its $CP$-conserving and violating parts $\vec F_0 + \delta \vec F$, 
\be
    \left({\partial\over\partial t}  + {\vec p\over E}\cdot\vec\nabla +
        \vec F_0\cdot \vec\nabla_p\right) \delta f_i = {\cal C}[\delta f_j] - \delta \vec F\cdot \vec\nabla_p f_0\, .
        \label{eq:boltz1}
\ee
Here $\vec\nabla_p$ is the gradient with respect to momenta,
${\cal C}$ is the collision term, linearized in the asymmetries $\delta f_j$, and $f_0$ is the unperturbed distribution function.  The last term
in Eq.\ (\ref{eq:boltz1}) leads to the $CP$-violating sources.

To derive the force, one approximately solves the Dirac equation for the
fermion in the background bubble wall fields, assuming that they are slowly varying compared to the de Broglie wavelength, $1/p$.  This gives a dispersion relation that varies locally within the wall, schematically of the form $p_z(z) = \sqrt{E^2 -p_\perp^2 + m^2(z)} \pm \delta p(z)$, where $\delta p(z)$ represents the $CP$-violating contributions, whose sign depends on the helicity of the particle (assumed in this example to be a fermion).  The $CP$-violating force is then given by 
\be
    \delta F = {d \over dt}\delta p = v_z {\partial \over \partial z}\delta p
\ee
in the rest frame of the wall.  $\delta p(z)$ depends upon the details of the particle's interactions and the background fields in the wall.
For example in the 2HDM as considered above, with spatially varying mass $m(z) = y_1 v_1 + y_2 v_2$, one finds that \cite{Fromme:2006cm,Cline:2011mm}\footnote{For brevity, we omit an additional term proportional to $|m^2|'(m^* m' - {m^*}' m)$ that tends to be numerically subdominant.}
\bea\label{eq:WKBsource}
   \delta  F &\cong& {1\over 2 E^2}(m^* m' - {m^*}' m)';\nn\\
    S_{\sss W\!K\!B} &=& { v_w\,\hat Q\over \gamma_w T^2}\, 2E^2 \delta F 
    \label{Swkb}
\eea
where $\hat Q\sim 0.15-0.3$ depends weakly on $v_w$,\footnote{We define $\hat Q = T^2 \gamma_w^2 Q^{(8,o)}_2$
in terms of the function $Q^{(8,o)}_2 = 3\tanh^{-1}(v_w)/(2\pi^2 v_w\gamma_w^2 T^2)$ (for massless particles) given in Ref.\ \cite{Cline:2020jre}.} 
and $\gamma_w$ accounts for the possibility of a relativistic bubble wall.  Eq.\ (\ref{Swkb})
has an additional spatial derivative compared to the VIA source term.  This will play a significant role in what follows.

The next step is to approximate the full Boltzmann equation by a set of
fluid equations for the particle density (or chemical potential $\mu_i$) and one higher moment, the velocity perturbation $u_i$, by integrating Eq.\ (\ref{eq:boltz1}) over spatial momenta, weighted by 1 and $p_z/E$, 
respectively.  Unlike the VIA formalism, this leads to first order fluid equations, with twice as many dependent variables per particle species.
Moreover the source term, which is the phase space average of $\vec p\cdot \delta\vec F/E$,
appears principally in the equation for $u_i$, due to the fact that 
$\vec\nabla_p \delta f_i$ nearly averages to zero, unless it is weighted by $p_z$.\footnote{In the rest frame of the wall, 
$\langle p/E\rangle \sim v_w$, so the source term 
in the equation for $\mu_i$ is not exactly zero, but it is suppressed by an additional power of $v_w$.}  The generic form of the WKB fluid equations is \cite{Joyce:1994fu,Cline:2000nw,Fromme:2006wx,Konstandin:2013caa,Cline:2020jre}
\be
    A_i\left({\mu_i\atop u_i}\right)' +(m^2_i)' B_i\left({\mu_i\atop u_i}\right)
    -{\cal C}_{i}^{\sss W\!K\!B} = \left(O(v_w S_{{\sss W\!K\!B},i})\atop S_{{\sss W\!K\!B},i}\right)
    \label{eq:fluid}
\ee
\smallskip

\noindent where $A_i$ and $B_i$ are dimensionless matrices that depend upon 
$m^2_i(z)/T^2$ and $v_w$.

To compare Eq.\ (\ref{eq:fluid}) with the VIA transport equation (\ref{HNeq}), one can approximately eliminate $u_i$ from the system (\ref{eq:fluid}) by taking an additional derivative, and ignoring the $z$-dependence of the matrices $A_i$ and $B_i$, resulting in a second-order diffusion equation for $\mu_i$ like Eq.\ (\ref{HNeq}).  In this way the WKB source term is transformed 
into its counterpart for use in the VIA diffusion equation, by replacing 
\be
    S_{\sss V\!I\!A} \to  {T^2\over 6}\left( {S'_{\sss W\!K\!B}\over \Gamma_{\rm tot}\gamma_w^2} + v_w {S_{\sss W\!K\!B}} \right)
    \label{Strans}
\ee
in Eq.\ (\ref{HNeq}), 
where $\Gamma_{\rm tot}$ is the total interaction rate of the species,
including elastic interactions, which is related to the diffusion constant by $D_i = \langle v^2\rangle/(3\Gamma_{i,\rm tot})$ \cite{Cline:2020jre}, and $\gamma_w = 1/\sqrt{1-v_w^2}$.  The factor $T^2/6$ arises from the relation between the particle asymmetry and chemical potential, $n = \mu T^2/6$ for a single chiral species.  
The second term in Eq.\ (\ref{Strans}) is negligible in practice.

{\it A priori}, this procedure might not seem justified for particles like quarks with relatively short diffusion lengths, since the distance scale for variations of $A_i, B_i$ is the bubble wall thickness $L_w$, which may not be small compared to the diffusion length.
Generally one might expect that taking two moments of the full Boltzmann equation should give more accurate results than taking just one.  Nevertheless it has been shown that, except for narrow 
$L_w$, the two formalisms give similar results, for equivalent source
terms \cite{Cline:2020jre}.

In contrast to VIA,  the WKB transport equations  and source term {\it have} been recovered from first principles, starting from the CTP formalism \cite{Kainulainen:2001cn}.  This rigorous derivation validiates the earlier constructions \cite{Cline:2000kb,Cline:2000nw,Huber:2000mg} based on
computing the semiclassical force from the Dirac equation and incorporating it into the Boltzmann equation.  It also shows that no sources arise at leading order in the derivative expansion from bosonic fields, unlike in the VIA.

\subsection{Collision rates}\label{sect:rates}
Over the years, estimates have been made for the rates of the various scattering processes that enter into the collision terms of the transport equations, with different values or conflicting normalizationa sometimes being adopted by different papers.  To facilitate comparison between the two formalisms, we list here the most recent values of the relevant rates that we adopt for the present study, and point out some cases where inconsistent values have been used in previous literature.

Again, one can compare the collision rates appearing in Eqs.\ (\ref{HNeq}) and (\ref{eq:fluid}) by reexpressing the latter as a second-order differential equation and substituting $\mu_i = 6n_i/(k_i T^2)$, where $k_i$ is the number of light degrees of freedom for fermions, or twice that number for bosons. Using that method, one finds that the respective collision rates used in the VIA and WKB transport equations (which are given explicitly in Appendix \ref{app:TransportEq}) are related by
\bea\label{eq:WKBrates}
\Gamma_{\sss Y,(q,\tau)}^{\sss V\!I\!A}&=&2k_{q,\tau}\Gamma_{\sss Y,(q,\tau)}^{\sss W\!K\!B}\nn\\
\Gamma_{\sss M,(q,\tau)}^{\sss V\!I\!A}&=&2k_{q,\tau}\Gamma_{\sss M,(q,\tau)}^{\sss W\!K\!B}\nn\\
\Gamma_{\sss W}^{\sss V\!I\!A}&=&k_{q}\Gamma_{\sss W}^{\sss W\!K\!B}\nn\\
\Gamma_{\sss H}^{\sss V\!I\!A}&=&k_{h}\Gamma_{\sss H}^{\sss W\!K\!B}\nn\\
\Gamma_{\sss S\!S}^{\sss V\!I\!A}&=&k_{q}\Gamma_{\sss S\!S}^{\sss W\!K\!B}\nn\\
\Gamma_{\sss W\!S}^{\sss V\!I\!A}&=&2k_q\Gamma_{\sss W\!S}^{\sss W\!K\!B}
\eea
where $k_q=3$, $k_\tau=1$ and $k_h=2$. Respectively, the quantities in (\ref{eq:WKBrates}) correspond to the Yukawa, helicity flip, $W$ boson scattering, Higgs damping, strong sphaleron and weak sphaleron rates. The values of the VIA rates adopted in this paper are\footnote{The $\tau$ Yukawa rate in Ref.\ \cite{Fuchs:2020pun} contains a contribution from the 4-point (${\bar l}_L \tau_R h g$)-interaction which should be replaced by the process (${\bar l}_L \tau_R h W$) since the leptons do not couple to the gluons. This can be done by replacing the factor $g_s^2$ by $3g'^2/8$.}
\bea
\Gamma_{\sss Y}^{\sss (q)} &\approx& 3.0\times10^{-2}\, T \quad \text{\cite{deVries:2017ncy}}\nn\\
\Gamma_{\sss M}^{\sss (q)} &\approx& {0.79m_t^2\over T} \quad\text{\cite{deVries:2017ncy}}\nn\\
\Gamma_{\sss Y}^{\sss (\tau)} &\approx& 4.4\times10^{-7}\, T\quad\text{\cite{Fuchs:2020pun}}\nn\\
\Gamma_{\sss M}^{\sss (\tau)}&\approx& {0.36m_\tau^2\over T} \quad\text{\cite{Fuchs:2020pun}}\nn\\ 
\Gamma_{\sss W} &\approx& {1\over 3D_h} \cong {T\over 60} \quad\text{\cite{Fromme:2006wx,Cline:1997vk}}\nn\\
\Gamma_{\sss H} &\approx&  {m_W^2\over 25T} \quad\text{\cite{Fromme:2006wx}}\nn\\
\Gamma_{\sss S\!S}&\approx& 8.7\times10^{-3}\, T \quad\text{\cite{Moore:2010jd}}\nn\\
\Gamma_{\sss W\!S} &\approx& 6.3\times10^{-6}\, T \quad\text{\cite{Bodeker:1999gx}}.
\eea

The strong sphaleron rate given here is different from what was often used in other VIA and WKB studies. It comes from  Ref.\ \cite{Moore:2010jd}, which used the normalization typically compatible with the WKB fluid equations, but differing by a factor of $k_q=3$ from the VIA's. 

We have updated the values of diffusion constants $D_q$,
$D_L$ and $D_R$ for quarks and  left-handed or right-handed leptons, respectively, using modern determinations from Refs.\ \cite{Arnold:2000dr,Arnold:2003zc} that include leading log contributions from the infrared-sensitive $t$-channel gauge boson exchanges:
\be
    D_q \cong {7.4\over T},\quad D_L\cong {90\over T},\quad
        D_R \cong {490\over T}\,.
        \label{eq:Dvals}
\ee
Previously, values estimated in Ref.\ \cite{Joyce:1994zt} were used in the literature: $D_q = 6/T$, $D_L = 100/T$, and $D_R = 300/T$.

\subsection{Weak sphaleron rate}
\label{sect:sphaleronrate}

For quantitative predictions, it is necessary to establish the correct value of the weak sphaleron 
rate $\Gamma_{\sss W\!S}$ appearing in the Boltzmann equation for baryon production,
\be
    \dot n_B = {n_f\over 2}\Gamma_{\sss W\!S} T^2\sum_i\mu_i\,,
    \label{eq:gsph}
\ee
where $n_f=3$ is the number of generations, and the sum is over chemical potentials of each
left-handed fermion doublet---not the the individual members of each doublet.  In this normalization,
$\Gamma_{\sss W\!S} T^3$ is the rate of Chern-Simons number density diffusion as measured in lattice computations
\cite{Bodeker:1999gx}, which are used to obtain the most precise determinations.  

However there is disagreement in the literature concerning the normalization of Eq.\ (\ref{eq:gsph}), leading to overestimation by factors of 2-4 in many papers.  In particular
Ref.\ \cite{Huet:1995sh} (see Eq.\ (3.23)) omits the factor of $1/2$ {\it and} counts individual members of doublets, giving an extraneous factor of 4.  Since the HN rate equations were adopted by VIA practitioners, this error (or part of it) has tended to propagate into later papers.  On the other hand the WKB Ref.\ \cite{Cline:2000nw} agrees with Eq.\ (\ref{eq:gsph}), and subsequent WKB studies tended to follow this (including Ref.\ \cite{Fromme:2006wx}, in contrast to the earlier works \cite{Joyce:1994bk,Cline:1997vk}, which counted both doublet members).

We are aware of one paper that has explicitly studied the relationship between the 
baryon violation rate appearing in Eq.\ (\ref{eq:gsph}), which is known as the linear response rate $\Gamma_\mu$, and the Chern-Simons diffusion rate that is directly measured on the lattice,
$\Gamma_d$.
Ref.\ \cite{Moore:1996qs} showed by detailed arguments that $\Gamma_\mu =  \Gamma_d/2$, which explains the factor of $1/2$ in Eq.\ (\ref{eq:gsph}), as well as the fact that each $\mu_i$ should refer to the doublet as a whole (with quarks weighted by a factor of 3 for color) and not
individual members of the doublet.  In the case where the two members have different chemical potentials, due to the nonvanishing Higgs VEV in the bubble wall, it is consistent to average over them.  

\section{Discrepancy between VIA and WKB predictions}
\label{sect:discrepancy}

As noted in Ref.\ \cite{Cline:2020jre}, the VIA typically predicts a BAU that is 
1-2 orders of magnitude larger than in the WKB approach. This discrepancy comes primarily from the difference in how the $CP$ asymmetry is generated by the moving wall, which is described by the source term. From Eqs.\ (\ref{eq:VIAsource},\ref{eq:WKBsource},\ref{Strans}), one can  show that the source terms appearing in the second-order diffusion equation in the two approaches are related at intermediate wall velocities by
\be
S_{\sss W\!K\!B}\sim \frac{D}{T} S''_{\sss V\!I\!A}\,.
\ee

\subsection{Analytic approximation}

A rough estimate of the BAU in either formalism can be made by solving the diffusion equation (\ref{HNeq}) with the Green's function method. It is convenient to first rewrite the $N$ second-order diffusion equations as a set of $2N$ first-order differential equation in matrix form,
\be\label{eq:matrixDiff}
q'(z)-{\bar\Gamma}q(z)={\bar S}(z),
\ee
\be\label{eq:greensource}
q(z) = \left( n(z)\atop n'(z)/T\right),\ \ {\bar S}(z) = \left( 0\atop \mathcal{D}^{-1}S(z)\right),
\ee
\be\label{eq:greenGamma}
{\bar \Gamma} = \left(\begin{array}{cc}
    0 & T \\
    \mathcal{D}^{-1}\Gamma & -v_w T\mathcal{D}^{-1}
\end{array}\right),
\ee
with $\mathcal{D}=T\,\mathrm{diag}(D_i)$, and the collision matrix $\Gamma$ is defined by $\sum_j\Gamma_{ij}n_j = \mathcal{C}_i[n]$. Neglecting all $z$ dependence in $\Gamma$, one can show that the Green's function of Eq.\ (\ref{eq:matrixDiff}) is 
\be\label{eq:green}
G_{ij}(z) = \left\lbrace \phantom{-}\sum_{\lambda_k<0} \rchi_{ik}\rchi_{kj}^{-1}e^{\lambda_k z},\ z>0 \atop -\sum_{\lambda_k\geq 0} \rchi_{ik}\rchi_{kj}^{-1}e^{\lambda_k z},\ z<0 \right.\,,
\ee
where $\lambda_i$ and $\rchi_{ij}$ are respectively the eigenvalues and  the matrix of eigenvectors of ${\bar\Gamma}$, which satisfy
\be\label{eq:eigvec}
\sum_j {\bar\Gamma}_{ij}\rchi_{jk}=\lambda_k\rchi_{ik}\,.
\ee

Although $\lambda_i$ and $\rchi_{ij}$ cannot be determined exactly, it is still possible to extract useful information from Eq.\ (\ref{eq:eigvec}). For a single eigenvector, it reads
\bea\label{eq:eigeq}
\lambda\left( \rchi_\uparrow \atop \rchi_\downarrow \right) &=& \left(\begin{array}{cc}
    0 & T \\
    \mathcal{D}^{-1}\Gamma & -v_w T\mathcal{D}^{-1}
\end{array}\right) \left( \rchi_\uparrow \atop \rchi_\downarrow \right) \nn \\
&=& \left( T\rchi_\downarrow \atop \mathcal{D}^{-1}\Gamma\rchi_\uparrow - v_w T\mathcal{D}^{-1}\rchi_\downarrow\right)\,. 
\eea
where $\rchi_\uparrow$ ($\rchi_\downarrow$) refers to the upper (lower) $N$ rows of
$\rchi$.
Interestingly, the upper half of (\ref{eq:eigeq}) implies that $\rchi_\downarrow$ and $\rchi_\uparrow$ are related by $\rchi_\uparrow = (\lambda/T)\rchi_\downarrow$, 
{\it i.e.,}
\be
\rchi_{i+N,j} = \frac{\lambda_j}{T}\rchi_{ij},\quad i=1,\cdots,N\,.
\ee
The lower half of (\ref{eq:eigeq}) can then be simplified to 
\be
\det(\lambda^2 + v_w\lambda T\mathcal{D}^{-1} - T\mathcal{D}^{-1}\Gamma)=0\,,
\ee
which can in principle be solved for $\lambda_i$. By neglecting consecutively one of the three terms in the determinant, one obtains a rough estimate for the positive and negative eigenvalues:
\bea
\lambda_i^{(+)} &\sim& \phantom{-}\min\left[ \sqrt{\frac{\gamma_i}{D_i}},\frac{\gamma_i}{v_w}\right] \nn \\ 
\lambda_i^{(-)} &\sim& -\max\left[ \sqrt{\frac{\gamma_i}{D_i}},\frac{v_w}{D_i}\right],
\eea
where $\gamma_i$ are the eigenvalues of $\Gamma$.

By integrating the Green's function (\ref{eq:green}) against the source term (\ref{eq:greensource}), one obtains the solution to the diffusion equation (\ref{eq:matrixDiff}),
\bea\label{eq:greensol}
q_i(z) &=& \int_{-\infty}^\infty dz'\, G_{ij}(z-z'){\bar S}_j(z') \\
&=& -\sum_{\lambda_j\geq0}\rchi_{ij}\int_z^\infty dz'\, [\rchi^{-1}{\bar S}(z')]_j e^{\lambda_j(z-z')} \nn \\
&& +\sum_{\lambda_j<0}\rchi_{ij}\int_{-\infty}^z dz'\, [\rchi^{-1}{\bar S}(z')]_j e^{\lambda_j(z-z')} \nn \\
&=& -\sum_{j}\rchi_{ij}\int_z^\infty dz'\, [\rchi^{-1}{\bar S}(z')]_j e^{\lambda_j(z-z')} \nn \\
&& +\sum_{\lambda_j<0}\rchi_{ij} e^{\lambda_j z}\int_{-\infty}^\infty dz'\, [\rchi^{-1}{\bar S}(z')]_j e^{-\lambda_j z'} \,.\nn
\eea
The motivation for rewriting (\ref{eq:greensol}) as the last two lines is that in this form, two qualitatively different contributions are isolated.  
 The first term decays in front of the wall\footnote{We take $z>0$ as the region in front of the wall} as fast as ${\bar S}(z)$, so it corresponds to the solution close to the wall, directly activated by the source. In contrast, the second term decays at large $z$ as $e^{-\lambda_i z}$, which has a much longer extent than $\bar S(z)$ if $\lambda_i\ll 1/L_w$. It corresponds to the diffusion of the $CP$ asymmetries in front of the wall.

One can determine how the BAU (denoted as $\eta_B$) scales with the fluid and wall properties by integrating the particle density, which corresponds to the upper $N$ components of equation (\ref{eq:greensol}). Integrating by parts, one gets
\bea\label{eq:BAU}
\eta_B &\sim& \frac{1}{T^2}\int_0^\infty dz\, n_i(z)\qquad (i=1,\cdots,N) \nn \\
&=& -\sum_j\frac{\rchi_{ij}}{T^2\lambda_j} \int_0^\infty dz[\rchi^{-1}{\bar S}(z)]_j(1-e^{-\lambda_j z}) \nn \\
&& -\sum_{\lambda_j<0}\frac{\rchi_{ij}}{T^2\lambda_j}\int_{-\infty}^\infty dz\, [\rchi^{-1}{\bar S}(z)]_j e^{-\lambda_j z} \,.
\eea
We now specialize to the case where $\lambda_i \ll 1/L_w$, allowing us to expand $\eta_B$ to leading order in $\lambda_i$. This is always valid for leptons due to their large diffusion lengths (\ref{eq:Dvals}), and it is a reasonable approximation for the quarks.
We consider first the contribution of the middle line, by Taylor expanding $-(1 - e^{-\lambda_i z}) = - \lambda_i z +\sfrac12 (\lambda_i z)^2+\dots$.  The term linear in $z$ vanishes because $\sum_j\chi_{ij}[\chi^{-1} \bar S_{\sss V\!I\!A}]_j =  [S_{\sss V\!I\!A}]_i$, which is zero for $i = 1,\dots,N$.
Therefore, the leading contribution comes from the next term in the expansion. Using Eqs.\ (\ref{eq:greensource},\ref{eq:greenGamma},\ref{eq:eigvec}), it gives
\be
\int_0^\infty dz\, z^2 \frac{S_{i}(z)}{2D_iT^2} \sim
    {S_0 L_w^2\over D T^2}\,.
    \label{eq:smallcont}
\ee
where $S_0/L_w$ is the magnitude of $S$ (taking into account that the VIA source contains one derivative with respect to $z$, to extract the correct dependence on $L_w$), and $L_w$ is its typical width.

Next we estimate the contribution from the last line of (\ref{eq:BAU}), which is expected to 
dominate in the regime of long diffusion lengths.
Considering the magnitude of the matrices $\rchi$ and $\rchi^{-1}$,
\be
\rchi\sim\left(\begin{array}{cc}
    \mathcal{O}(1) & \mathcal{O}(1) \\
    \mathcal{O}(\lambda/T) & \mathcal{O}(\lambda/T)
\end{array}\right), \
\rchi^{-1}\sim\left(\begin{array}{cc}
    \mathcal{O}(1) & \mathcal{O}(T/\lambda) \\
    \mathcal{O}(1) & \mathcal{O}(T/\lambda)
\end{array}\right)\,,
\ee
we find that
\bea
\sum_{\lambda_j<0}\frac{\rchi_{ij}}{\lambda_j}[\rchi^{-1}{\bar S}]_j &\sim& A_i\left(\begin{array}{cc}
    1/\lambda & 1/\lambda \\
    1/T & 1/T
\end{array}\right)
\left(\begin{array}{cc}
    1 & T/\lambda \\
    1 & T/\lambda
\end{array}\right) 
\left( 0 \atop \mathcal{D}^{-1}S\right) \nn \\
&\sim& A_i\left(T\mathcal{D}^{-1} S/\lambda^2 \atop \mathcal{D}^{-1} S/\lambda\right),
\eea
where $A_i\sim\mathcal{O}(1)$ depends only weakly on $v_w$, $\Gamma$ and $D$. Therefore, the leading order contribution to the BAU in the VIA method is given by
\bea\label{eq:BAUVIA}
\eta_B^{\sss V\!I\!A}&\sim& \int_0^\infty dz\, z^2 \frac{S_{\sss V\!I\!A}(z)}{2DT^2} - A\int_{-\infty}^\infty dz\,\frac{S_{\sss V\!I\!A}(z)}{\lambda^2DT^2}e^{-\lambda z} \nn \\
&\sim&\frac{S_0 }{\lambda^2D T^2}\,,
\eea
which is evidently dominated by the tail of the solution coming from the diffusion of the particles in the plasma, and is larger than the contribution (\ref{eq:smallcont}) by the factor $(L_w\lambda)^{-2}$.

To obtain the BAU from the WKB method, one only has to replace the source term in the last expression by $S_{\sss W\!K\!B}\sim DS''_{\sss V\!I\!A}/T$ and integrate by parts twice:
\bea\label{eq:BAUWKB}
\eta_B^{\sss W\!K\!B} &\sim& \int_0^\infty dz\, \frac{S_{\sss V\!I\!A}(z)}{T^3} - A\int_{-\infty}^\infty dz\,\frac{S_{\sss V\!I\!A}(z)}{T^3} \nn \\
&\sim& {S_0\over T^3} 
\eea
The additional two derivatives in the WKB source have suppressed the contribution from the tail of the solution by a factor of $\lambda^2$. The region close to the wall directly activated by the source is now as important as the tail. Remarkably, apart from the small variations of $A$, this expression does not depend on the diffusion of the particles in the plasma or the wall velocity. This implies that, to leading order, only the source term itself is important in the WKB formalism.

From Eqs.\ (\ref{eq:BAUVIA},\ref{eq:BAUWKB}), one can estimate the discrepancy between the two approaches to be 
\be\label{eq:discrepancy}
\left\vert\frac{\eta_B^{\sss W\!K\!B}}{\eta_B^{\sss V\!I\!A}}\right\vert\sim \frac{\lambda^2D}{T} \sim {\rm max}\left[{\gamma\over T},\,{v_w^2\over DT}\right]\,.
\ee
The smallness of the ratio is due to the largest contribution in the VIA coming from the tail of the solution, which has a typical amplitude and length of $1/\lambda$. In contrast, the specific shape of the WKB source reduces the amplitude of the tail by $\lambda^2$, without changing its length. Therefore in the limit where the particles do not interact at all, the length of the tail diverges while its amplitude goes to zero, in such a way that its contribution to the BAU is approximately independent of $\lambda$. 

\begin{figure*}[t]
\begin{center}
\centerline{
\includegraphics[width=0.48\textwidth]{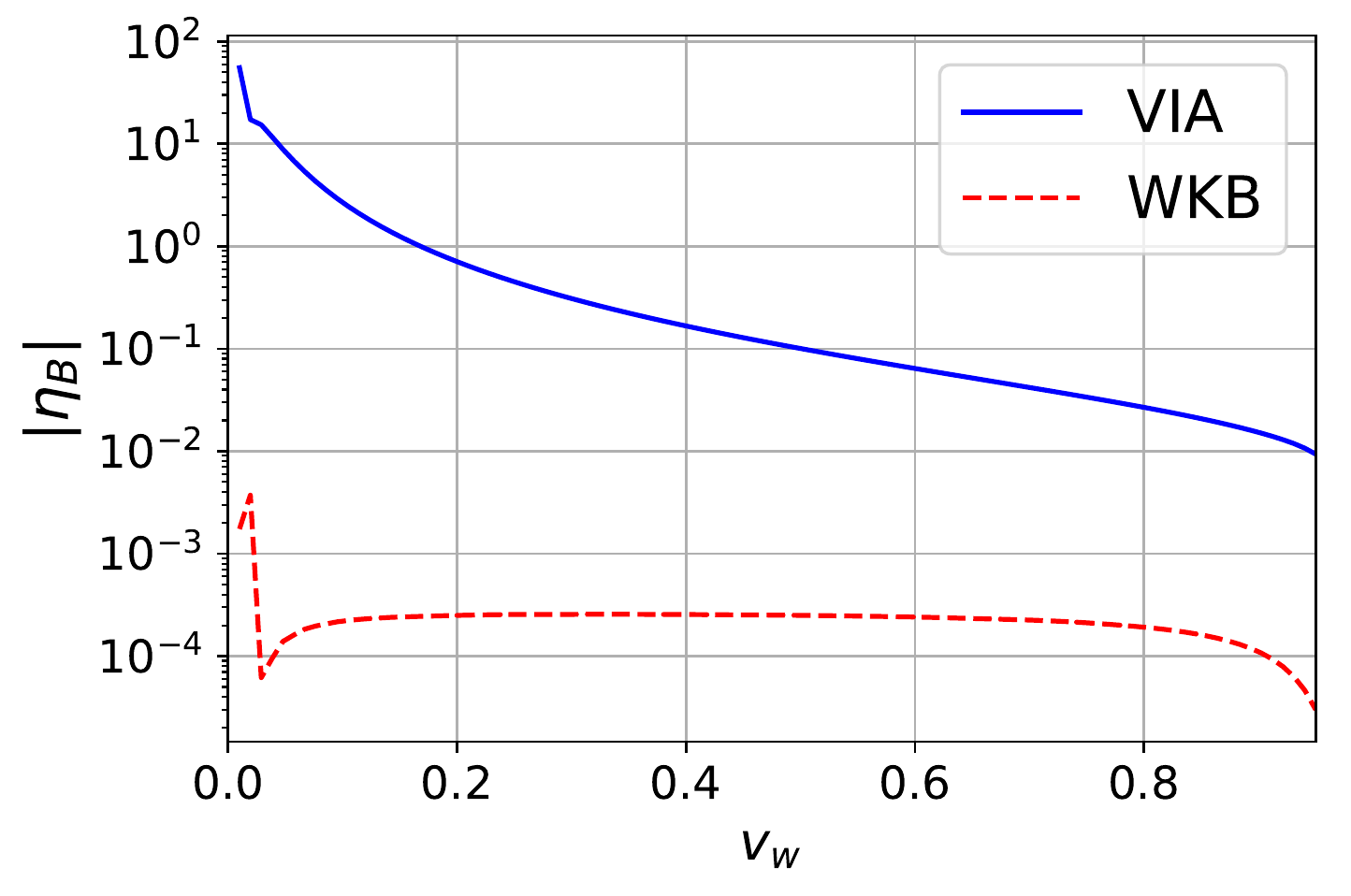}
\hspace{0.04\textwidth}\includegraphics[width=0.48\textwidth]{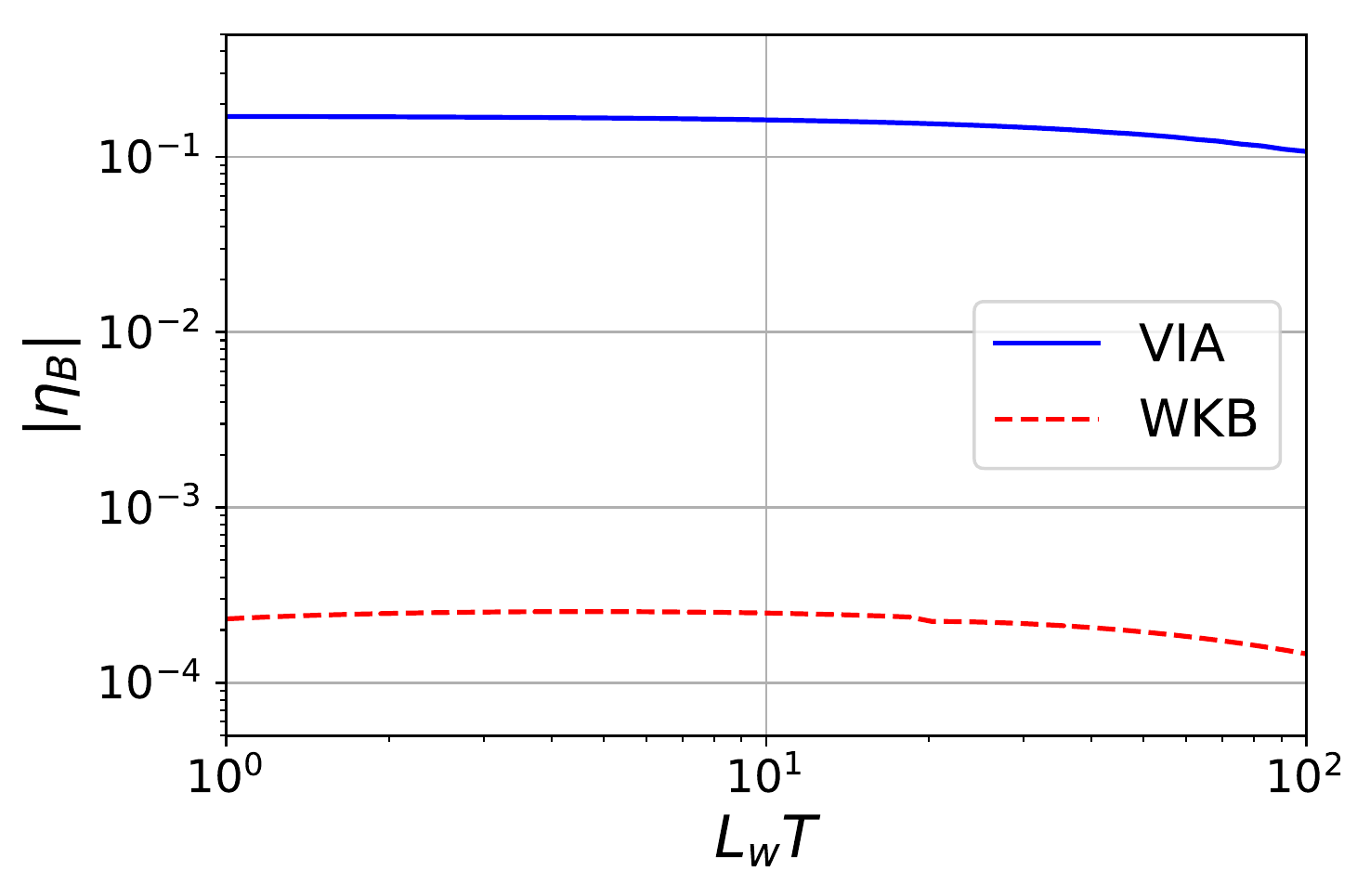}}
 \caption{Baryon asymmetry (in units of observed asymmetry) predicted by  numerically solving the WKB and VIA transport equations for a $\tau$ lepton source as a function of the wall velocity $v_w$ (left) and wall thickness $L_w$ (right). The left plot was computed with $L_w=5/T$ and the one at right with $v_w=0.4$.} 
 \label{fig:leptonsBAU}
\end{center} 
\end{figure*}

\subsection{Application to quark mass sources}

We illustrate the previous estimates for the case of a $CP$-violating quark mass term of the
form $m(z) = |m(z)|e^{i\theta(z)}$, which typically arises in 2HDMs or singlet extensions of the SM.  
Consider first a top quark source. For quarks, the eigenvalues of $\Gamma$ and the diffusion rates are of order $\gamma\sim0.1\,T$ and $D\sim10/T$, respectively. The transport equations' eigenvalues are therefore $\lambda\sim\sqrt{\gamma/D}\sim0.1\,T$ which leads to the WKB method predicting a BAU roughly 10 times smaller than the VIA, in agreement with the results of Ref.\ \cite{Cline:2020jre}.

One  can rescale the results of Ref.\ \cite{Cline:2020jre} and Eqs.\ (\ref{eq:BAUVIA},\ref{eq:BAUWKB}) to estimate the BAU for analogous sources from lighter
quark flavors.
 Even though the network of diffusion equations may differ slightly depending on the flavor of the quark source, the collision terms are mostly dominated by the strong sphaleron rate, which is the same for all flavors. Therefore the eigenvalues $\lambda_i$ depend only weakly on the details of the diffusion equations, as long as they describe quarks. One concludes that in both the VIA and WKB methods, the BAU should scale like $S_0$, which is  proportional to the Yukawa coupling squared. 

Recently Ref.\ \cite{Shapira:2021mmy} considered the $c$ quark contribution to the BAU
within the VIA, finding that it could account for at most 1\% of the total BAU.  Using the previous estimates, we predict $\eta_B^{\sss V\!I\!A}\sim10^{-3}$ in units of the observed asymmetry.   The VIA predictions of  \cite{Shapira:2021mmy} become roughly consistent with this when we take into account the correction 
factors $1/3\times 1/2$ arising from the strong and weak sphaleron rates discussed in sections 
\ref{sect:rates}-\ref{sect:sphaleronrate}, under which the BAU scales as  $\eta_B^{\sss V\!I\!A}\sim \Gamma_{\sss W\!S}/\Gamma_{\sss S\!S}$.   This turns out to be an order of magnitude larger than the WKB prediction, estimated
by rescaling the top quark results of Ref.\ \cite{Cline:2020jre} by $(y_c/y_t)^2\sim 10^{-4}$; hence we find that $\eta_B^{\sss W\!K\!B}\sim10^{-4}$ for the $c$
quark source
in the WKB method.

\begin{figure*}[t]
    \centering
    \includegraphics[width=0.47\textwidth]{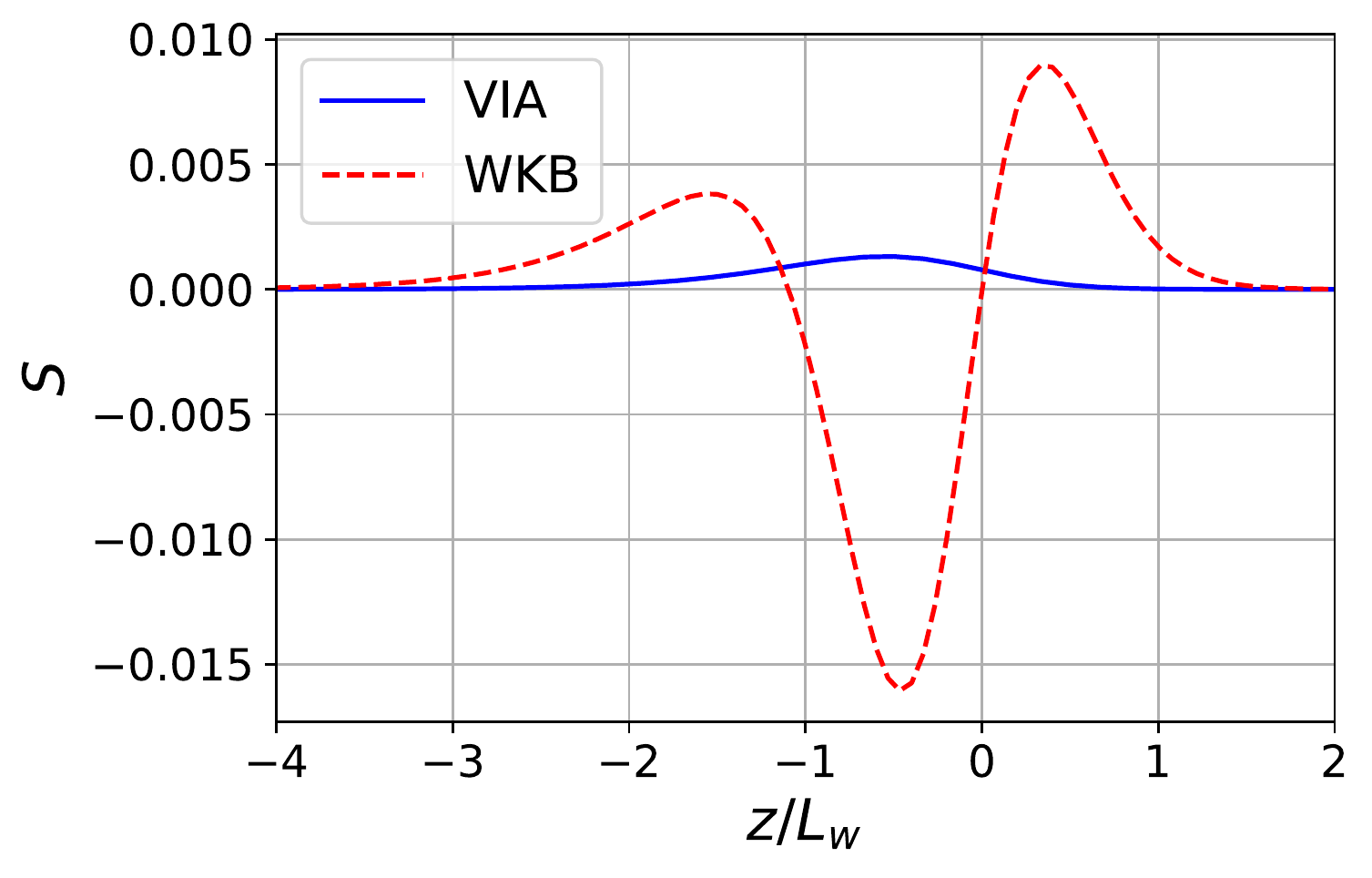}\hspace{0.04\textwidth}\includegraphics[width=0.45\textwidth]{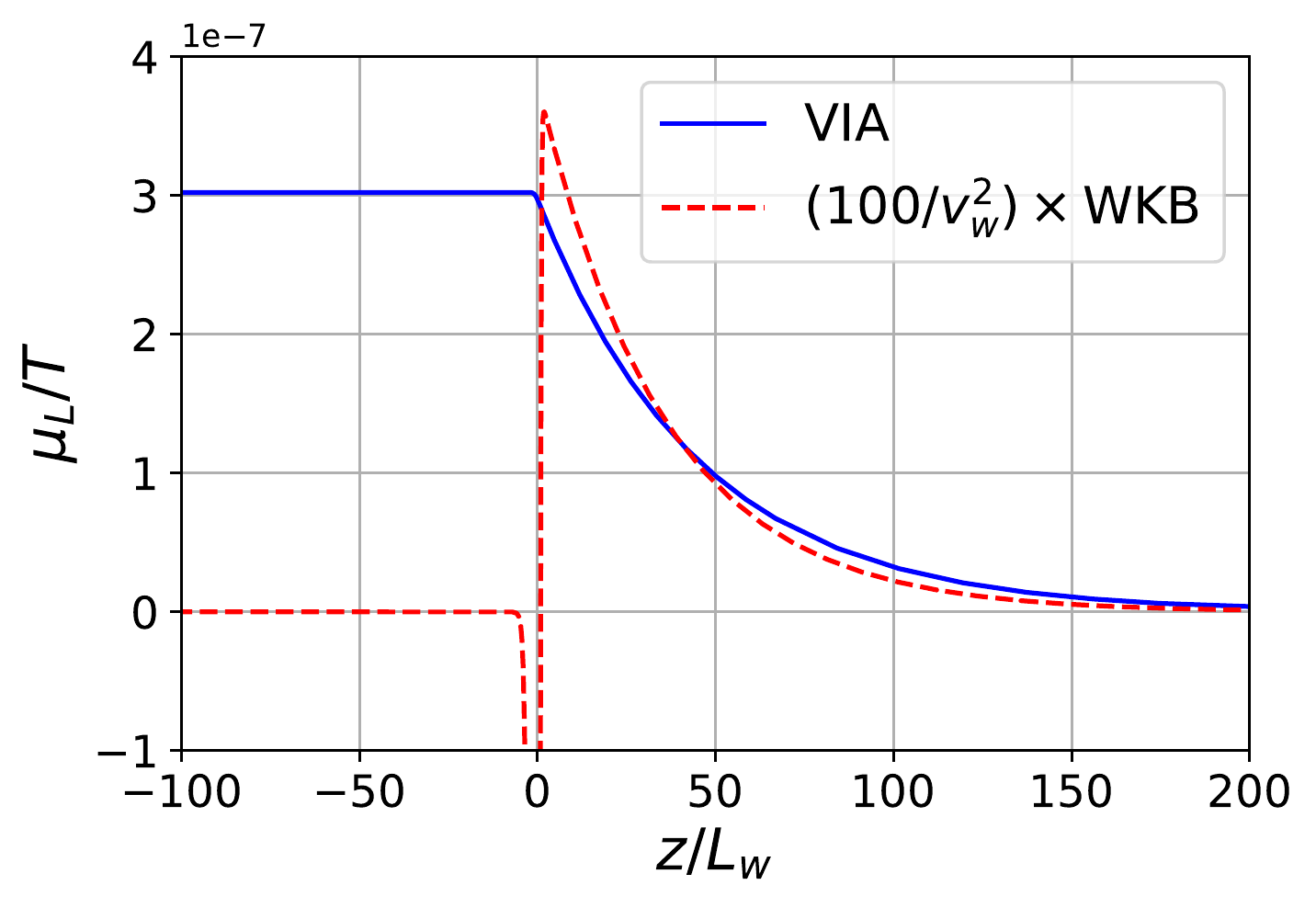}
    \caption{Left: Source term in the VIA and WKB formalisms. Right: Chemical potential of the LH leptons obtained by solving numerically the VIA and WKB transport equations for the $\tau$ lepton source. The WKB solution is magnified by 100$/v_w^2$ to make it more visible relative to  the VIA solution (see Eq.\ \ref{eq:estimateLepton})). We used $v_w=0.4$ and $L_w=5/T$.}
    \label{fig:muL}
\end{figure*}

\section{EWBG from a $\tau$ lepton source}
\label{sect:tau}

In this section we will quantitatively compare the predictions for EWBG sourced by
a $CP$-violating contribution to the $\tau$
lepton mass, in the case where it couples to both the Higgs boson $h$ and to a singlet scalar $s$ field via a dimension-5 operator:
\bea
    {\cal L}_{\tau} &=& {y_\tau\over \sqrt{2}} \bar \tau_L h \left( 1 + 
        {i s\over\Lambda}\right)\tau_R + {\rm H.c.}\nn\\
        &\equiv& \bar \tau_L m(z) \bar \tau_R + {\rm H.c.}\,.
        \label{taumass}
\eea
It is assumed that $\langle s\rangle =0$ at low temperatures, and the singlet profile is only present in the bubble wall during the EWPT.
For simplicity we have chosen the maximum possible $CP$-violating phase between the two contributions, and we parametrize the wall profiles as
\be
    h(z) = {h_0\over 2}\left(1 - \tanh {z\over L_w}\right), \quad
     s(z) = {s_0\over 2}\left(1 + \tanh {z\over L_w}\right)\,,
     \label{eq:profile}
\ee
consistently with the assumption that $\langle s\rangle = 0$ in the electroweak symmetry
breaking vacuum.

The VIA source term is found to be \cite{Xie:2020wzn}
\bea
    S_{\sss V\!I\!A} &=& J_\tau{v_w T\over \pi^2}{\rm Im} (m'm^*)
        = {v_w T y_\tau^2 J_\tau\over 2\pi^2 \Lambda}\, h^2 s'\,,
        \label{Svtau}
\eea
where  the numerical factor $J_\tau\cong 0.55$ arises from the thermal phase space integral.  The corresponding WKB source can be inferred by analogy to the top quark case \cite{Cline:2020jre},
\be
    S_{\sss W\!K\!B} = {v_w y_\tau^2\,\hat Q\over 2\Lambda\, \gamma_w\,T^2}
    \left( h^2 s'\right)'\,.
\ee
As mentioned before, the WKB source has an additional derivative with respect to $z$ compared to its VIA counterpart, and from Eq.\ (\ref{Strans}), it will acquire yet one more derivative  (or power of $v_w$ suppression, arising from the $O(v_w S_{\sss W\!K\!B})$ term in Eq.\ (\ref{eq:fluid})), when converting from first-order to second-order fluid equations.

The detailed fluid equation networks in the respective formalisms are presented in Appendix \ref{app:TransportEq}. To simplify them, one can use the fact that the lepton subsystem is dominated by the $W$ interaction rate $\Gamma_W$, which tends to equalize the $\tau_L$ and $\nu_{\tau,L}$ densities (or chemical potentials). Therefore, one need not keep track of both particles separately and can instead describe the left-handed leptons by a single density $n_{l_L}=n_{\tau_L}+n_{\nu_{\tau,L}}$.  This works because of the
hierarchy  $\Gamma_W\gg \Gamma_M^{(\tau)}$ relative to the small relaxation rate $\Gamma_M^{(\tau)}$ from the mass-induced helicity-flipping interactions for leptons. 
On the other hand, for third-generation quarks, the large Yukawa coupling  leads to  $\Gamma_M^{(t)}\gg\Gamma_W$, making it necessary to track $b_L$ and $t_L$ separately. 

Before carrying out the quantitative numerical comparison, 
one can use the results of the previous section to anticipate the dependence of the BAU on the different parameters of the diffusion equations and the magnitude of the difference between the VIA and WKB methods. For leptons, the collision and diffusion rates are respectively of order $\Gamma\sim 10^{-6}T$ and $D\sim100/T$, so the eigenvalues are $|\lambda|/T\sim\max\left[10^{-4},v_w/100\right]$. From Eqs.\ (\ref{eq:BAUVIA},\ref{eq:BAUWKB}), the BAU in the two approaches scales as
\bea\label{eq:estimateLepton}
\eta_B^{\sss V\!I\!A} &\sim& \frac{y_\tau^2h_0^2 s_0}{\Lambda T}\min\left[\frac{1}{\Gamma},\frac{D}{v_w^2}\right]\sim \frac{y_\tau^2 h_0^2 s_0}{\Lambda T^2}\min\left[10^6,\frac{100}{v_w^2}\right]\,, \nn \\
\eta_B^{\sss W\!K\!B} &\sim& \frac{y_\tau^2 h_0^2 s_0}{\Lambda T^2}\,.
\eea
Comparing to the top quark source, the BAU in this model is suppressed by a factor of $(y_\tau/y_t)^2\sim 10^{-4}$. However, in the VIA, it is also enhanced by the small collision rates and large diffusion constants, which is enough to compensate for the small $y_\tau$ if $v_w\lesssim 0.1$. This is not the case for WKB, where the contribution from the tail of the solution does not depend on $\Gamma$ or $D$ to leading order. This leads to disagreement between the two methods by  factors as large as $10^6$.

The results of numerically solving the 
diffusion equations are  shown in Fig.\ \ref{fig:leptonsBAU}. The agreement between the analytic estimate (\ref{eq:estimateLepton}) and the numerical solutions is good, especially at intermediate velocities. Since the eigenvalues for this system are small, the leading term in the  expansion in $\lambda$ is dominant. As expected, $\eta_B^{\sss W\!K\!B}$ is nearly independent of the wall velocity, while $\eta_B^{\sss V\!I\!A} \sim 1/v_w^2$ for 
$v_w\gtrsim0.01$. In either approach the BAU depends only weakly on the wall thickness. For $0.1\lesssim v_w\lesssim0.8$, the ratio of the WKB and VIA curves is well approximated by Eq.\ (\ref{eq:discrepancy}), which gives $\sim100/v_w^2$.

The main result of this section is that the disagreement between the VIA and WKB methods is exacerbated by the leptons' weak interactions. {The particular shape of the WKB source leads to a suppression of the $CP$-violating chemical potentials $\mu_i$ in front of the wall, that
source the baryon asymmetry in Eq.\ (\ref{eq:gsph}).
It results from  the fact that $S_{\sss W\!K\!B}$ is a second derivative, which causes the solutions to vanish in the limit $v_w\rightarrow 0$ and $\Gamma\rightarrow0$, or equivalently $\lambda\rightarrow0$. The resulting suppression can be
understood from integrating by parts (twice) the WKB
source term in Eq.\ (\ref{eq:BAUWKB}):
\be
    \int dz\, S'' e^{-\lambda z} \sim \lambda^2 S_0 L_w\,.
\ee}

To illustrate, we show the source term and the chemical potential of LH leptons for VIA and WKB in Fig.\ \ref{fig:muL}. Even though the magnitude of the WKB source is larger than the VIA's, we see that the solution in front of the wall ($z>0$) is smaller by a factor of $\lambda^2/D\sim v_w^2/100\sim 10^{-3}$. Unlike the VIA source which has the same sign everywhere, the WKB source changes sign several times inside the wall in such a way that the chemical potential gets nearly cancelling contributions.  Hence the BAU turns out to be smaller in WKB, despite the fact that $S_{\sss W\!K\!B}$ is larger
in magnitude than $S_{\sss V\!I\!A}$.

\begin{figure*}[t]
\begin{center}
\centerline{
\includegraphics[width=0.48\textwidth]{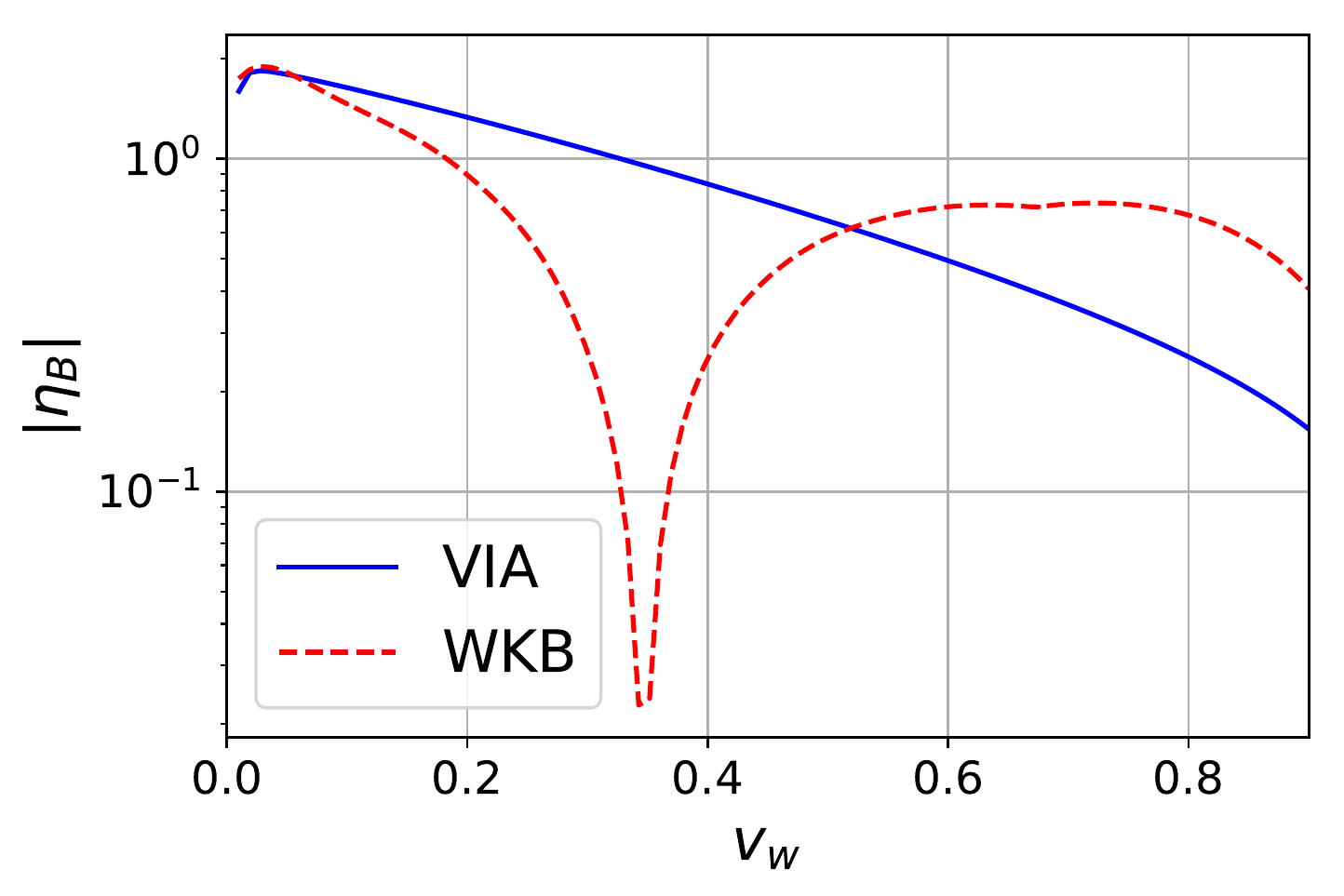}
\hspace{0.04\textwidth}\includegraphics[width=0.48\textwidth]{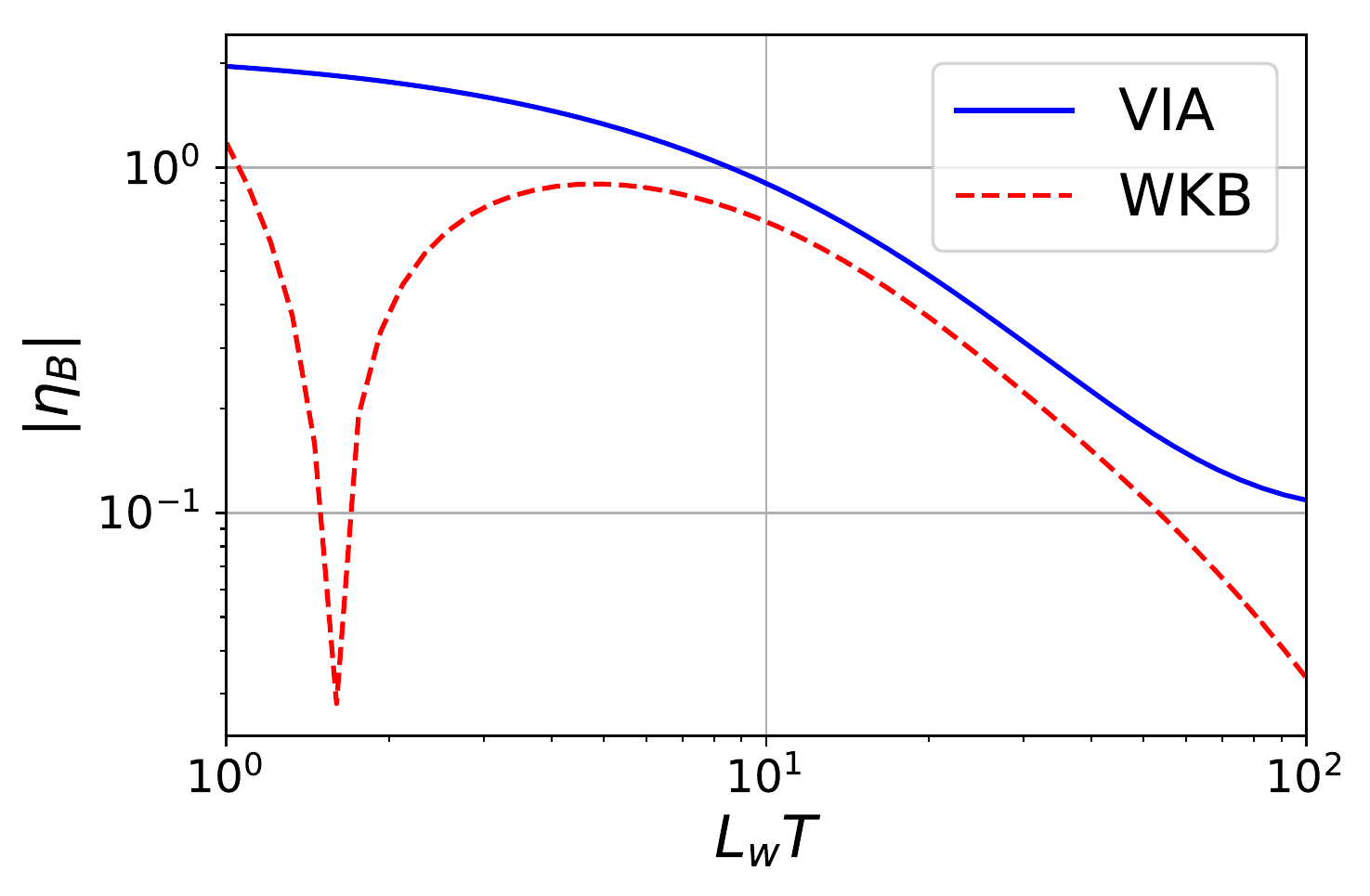}}
 \caption{Similar to Fig.\ \ref{fig:leptonsBAU}, but for a $CP$-violating source from $c$-$t$ quark mixing. The second plot was computed with $v_w=0.2$.} 
 \label{fig:topFCBAU}
\end{center} 
\end{figure*}

\section{EWBG from quark mass mixing}
\label{sect:mixing}

Models with $CP$ violation in an off-diagonal quark mass matrix element are similar to the diagonal case, Eq.\  (\ref{taumass}), in that an extra field is presumed to be varying inside the bubble wall, leading to a spatially dependent $CP$-violating phase for the mass eigenvalues.  This can arise in 2HDMs if the ratio of the VEVs $\tan\beta = \langle h_2\rangle / \langle h_1\rangle$ depends upon $z$ inside the wall.

\subsection{$c$-$t$ mixing}

For the case of $c$-$t$ quark mixing, the mass term is
$\bar q_L M q_R + {\rm H.c.}$ with mass matrix taken to have the form
\be\label{eq:massMatrix}
    M = {1\over\sqrt{2}}\left(\left({y_{cc}\atop 0}\,{y_{ct}\atop y_{tt}}
        \right) h_1 + \left({y_{cc}\atop 0}\ {e^{i\theta} y_{ct}\atop y_{tt}}
        \right)h_2\right)\,,
\ee
in an example where $\tan\beta=1$ at $T=0$, and $\theta$ is the $CP$-violating phase.  
We take the two Higgs field profiles in the bubble wall to be
\bea
    h_1(z) &=& \frac{v_0}{2\sqrt{2}}(1-\tanh(z/L_w)),\nn \\
    h_2(z) &=& \frac{v_0}{2\sqrt{2}}(1-\tanh(z/L_w+\Delta\beta)).
\eea
For a small offset $\Delta\beta$ between the two fields, this configuration is equivalent to the following $\beta(z)$ and $v(z)=\sqrt{h_1^2+h_2^2}$ profiles:
\bea
\beta(z) &=& \frac{\pi}{4} - \frac{\Delta\beta}{2}(1+\tanh(z/L_w)), \nn \\
v(z) &=& \frac{v_0}{2}(1-\tanh(z/L_w)),
\eea
where $\Delta\beta$ can manifestly be interpreted as the maximum variation of $\beta$. We take $\Delta\beta=0.015$ for the rest of this section, which is the value that was adopted by Refs.\ \cite{Fuyuto:2017ewj,Modak:2018csw}.
This is near the top of the range $\Delta\beta\lesssim 0.02$ found for the MSSM in Ref.\ \cite{Moreno:1998bq}.

{The parameters in Eq.\ (\ref{eq:massMatrix}) are constrained by the requirement that the eigenvalues of $(M^\dagger M)^{1/2}$ correspond to the observed quark masses.
Approximating $m_c\ll m_t$ and taking $\tan\beta=1$, one obtains
\bea\label{eq:yukawaConst}
y_{cc} &\cong& 0, \nn \\
(y_t^{\sss S\!M})^2 &=& 2y_{tt}^2+(1+\cos\theta)y_{ct}^2.
\eea
where $y_t^{\sss S\!M}\cong 1$ is the standard model value of the top quark Yukawa coupling.
A favorable choice for EWBG is to 
set $y_{tt}=0$, $y_{ct}=y_t^{\sss S\!M}$ and $\theta=\pi/2$ (maximal $CP$ violation), which are the values we adopt here.
}

The source term in the VIA is \cite{Fuyuto:2017ewj}
\bea
    S_{{\sss V\!I\!A},t_L} &=& -S_{{\sss V\!I\!A},c_R} = {v_w \over\pi^2}\,y_{ct}^2\,J_q\,T\,\sin\theta\,(h_1'h_2 - h_2' h_1)\,, \nn \\
    S_{{\sss V\!I\!A},t_R} &=& 0\,,
    \label{Svevq}
\eea
where $J_q\sim 0.4$ is a similar numerical factor as in Eq.\ (\ref{Svtau}).  

The corresponding WKB source can be derived in analogy to that for charginos in the minimal supersymmetric standard model \cite{Cline:2017qpe}.  One starts by constructing the unitary transformation $U$ such that $U M M^\dagger U^\dagger$ is locally diagonalized, and from it the matrix $A = {\rm Im}[U M\partial_z M^{-1} U^\dagger]$.  The
$CP$-violating force acting on the heavy eigenstate is given by
$2 E^2\delta F = A_{22}' = (M_{12}^*M_{12}'-M_{12} {M_{12}^*}')/2$, using the approximation 
$y_{cc}\cong 0$.  This leads to the WKB source
\bea
    S_{{\sss W\!K\!B},t_L} &=& -S_{{\sss W\!K\!B},t_R} = {v_w y_{ct}^2 \hat Q\over 2\gamma_w T^2}
    \,\sin\theta\,(h_1'h_2 - h_2' h_1)' \,,\nn \\
    S_{{\sss W\!K\!B},c_R} &=& 0\,.
    \label{Swkbq}
\eea
In this class of models, not only does the WKB source have an additional derivative compared to the VIA's, but different particle species are sourced. In the semiclassical formalism, the source term is predicted to be proportional to the particle's mass, which vanishes for the charm quark (in the present approximation). In the VIA, the $c_R$ quark gets  sourced, but not the $t_R$. This constitutes another qualitative difference between the two methods.

The explicit networks of equations in the VIA and WKB formalisms can be found in Appendix \ref{app:TransportEq}. In contrast to leptons, the third family of quarks is not dominated by the $W$ interaction rate. We therefore need to keep track of $t_L$ and $b_L$ separately.

The predicted BAU is shown in Fig.\ \ref{fig:topFCBAU}. The agreement between the two methods is much better than for the lepton case, except for wall speeds near $v_w\cong 0.45$ and the wall width near $L_w\cong1.5/T$, where the WKB result crosses zero.  This extra $v_w$-dependence in the WKB method can be understood from 
Eq.\ (\ref{eq:BAUWKB}), where the two terms with opposite signs and similar magnitude can accidentally cancel due to the $v_w$-dependence of the factor $A$.  

We find that the apparent better agreement between WKB and VIA in the quark mixing model is also accidental, coming 
about because the different structure of the two sources (in terms of how they appear in the transport equations)
compensates for the difference in orders of derivatives
between the two.  
The rough estimates derived in Section \ref{sect:discrepancy} are less accurate than for lepton sources in this model: unlike those predictions, the full numerical solutions give similar 
dependences on the wall velocity (apart from the zero-crossing in WKB) and thickness  for both formalisms. The reduced accuracy of the analytic estimates
is to be expected  since for quarks, the transport equations' eigenvalues are of order $1/L$, so  that the higher order terms in the eigenvalue expansion that were neglected can become important. Both approaches hence lead to the conclusion that at small $v_w$ and $L$, a $CP$-violating source from $c$-$t$ quark mass mixing can be strong enough to generate the observed BAU.

\subsection{$b$-$s$ mixing}
Completely analogous expressions to (\ref{Svevq}-\ref{Swkbq}) apply for the case of $b$-$s$ quark mixing, where EWBG was studied using the VIA in Ref.\ \cite{Liu:2011jh,Modak:2018csw,Modak:2021vre}.\footnote{Ref.\ \cite{Liu:2011jh} assumed an optimistically large value
$\Delta\beta = 0.05$, while Ref.\ \cite{Modak:2018csw} took
$\Delta\beta = 0.015$ for the change of $h_2/h_1$ in the
wall.}
We verified that the main difference comes from the rescaling of the source by a factor $(y_{sb}/y_{ct})^2$, which according to Eq.\ (\ref{eq:yukawaConst}) is of order $(y_b^{\sss S\!M}/y_t^{\sss S\!M})^2\sim\mathcal{O}(10^{-3})$. Otherwise the two systems behave similarly. Comparing with Fig.\ \ref{fig:topFCBAU}, this indicates that EWBG from $b$-$s$ mixing would be two or three orders of magnitude too small to explain the observed BAU. This seems to be in contradiction with the results of Ref.\ \cite{Liu:2011jh,Modak:2018csw}, which found regions of the parameter space with efficient EWBG. We have traced the discrepancy to the accumulation of several approximations and errors made in Ref.\ \cite{Liu:2011jh}, which have been corrected here. 

As discussed in Section \ref{sect:rates}, the strong sphaleron rate used in previous VIA studies was too small by a factor of 3, which had the effect of increasing the BAU by the same factor. A factor of 1/4 (see Sect.\ \ref{sect:sphaleronrate}) was  omitted in the sphaleron rate equation in Ref.\ \cite{Liu:2011jh}. Several approximations were made in order to solve the diffusion equations, some being quite inaccurate. For example, the spatial variation of $\beta'$ was neglected and the source term was assumed to vanish in front of the wall. We find that these approximations lead to overestimates by factors  of about 10 and 5, respectively in the BAU. Moreover, Ref.\  
 \cite{Liu:2011jh} found a BAU scaling linearly with the wall velocity which  contradicts the behavior of the full numerical solutions shown in Fig.\ \ref{fig:topFCBAU}. The same methodology was followed in Refs.\ \cite{Modak:2018csw,Modak:2021vre} which resulted in a
similar overestimate.

\section{Summary and conclusions}
\label{sect:concl}

In this work we carried out a detailed comparison of the VEV-insertion approximation  and the semiclassical  method to compute the baryon asymmetry resulting from a first-order electroweak phase transition, applying the two methods to different models involving $CP$-violating sources from light fermions. Both approaches agree that a source term from $c$-$t$ quark mixing can be consistent with
a  BAU of the observed value, although this agreement is accidental, due to the different structures
of the source term in flavor space between the two formalisms.

For the case of a $b$-$s$ mixing source, we have redone
the calculation within the VIA formalism to correct for
some errors and inconsistencies in previous literature.
The coincidental agreement with WKB then persists in this
model, giving a suppression of the BAU by a factor of
 $(y_b^{\sss S\!M}/y_t^{\sss S\!M})^2\sim10^{-3}$ relative to a flavor-diagonal top quark source.  This is too small to generate the observed BAU, by three orders of magnitude.
  For a source term from the $\tau$ lepton, the predictions of VIA versus WKB
are found to disagree by a factor of $10^3$--$10^5$, depending upon assumptions about the wall
speed and thickness.

To understand the origin of this discrepancy, we derived an analytic estimate of the solutions of the transport equations using the Green's function method. By expanding to leading order in the system's eigenvalues $\lambda_i$, which are small for weakly interacting particles, we found that the ratio between the VIA and the WKB  predictions is proportional to $\lambda_i^2\ll 1$. This disparity comes from the contribution from the tail of the solution, which has a typical length of $1/\lambda$. In the WKB, the specific shape of the source reduces the amplitude of the tail by $\lambda^2$ compared to the VIA's, resulting in a smaller BAU. These estimates were confirmed by the full numerical solutions of the transport equations, and they agree with results of Ref.\ \cite{Cline:2020jre}, which found a discrepancy of order $10^1$--$10^2$ between the VIA and WKB methods for a top quark source.

This outcome calls into question the viability of 
EWBG from a $\tau$ {lepton} source.  According to the VIA, the BAU production is sufficient at small $v_w$,
while the WKB method does not admit a large enough BAU at any wall speed. This highlights the need for a better understanding of the generation of the $CP$ asymmetry during a first-order phase transition. Some progress in that direction was recently made in Ref.\ \cite{Postma:2021zux}, where it was shown that the VIA for leptons is only valid for very thick walls $L_w\gg 30/T$, far beyond the typical thickness of an electroweak bubble wall. This seems to indicate that the large BAU obtained with the VIA should not be trusted in this model, and that a $CP$-violating source from leptons is insufficient for electroweak baryogenesis.

We have also pointed out and corrected several inconsistencies in the values of strong and weak sphaleron rates used in previous literature, and updated values of
diffusion constants, that may be useful for future studies of EWBG.

{\bf Acknowledgements.}  We thank K.\ Kainulainen , T.\ Konstandin, M.\ Postma
and K.-P.\ Xie for valuable correspondence, and G.\ Moore for enlightening discussions about the weak sphaleron rate.
This work was supported by the Natural Sciences and Engineering Research Council (NSERC) of Canada. and the Fonds de recherche du Québec Nature et technologies (FRNQT).

\begin{appendix}

\section{Transport equations}
\label{app:TransportEq}
For the convenience of the reader, we here write explicitly the general form of the transport equation used in the VIA and WKB methods.

The fluid equation network in the VIA formalism has the form
\be
    D_i n_i''+v_w n_i'-{\cal C}_{i}^{\sss V\!I\!A}[n_j]=S_{{\sss V\!I\!A},i},
\ee
for $i=l_L,\tau_R,t_L,b_L,t_R,\delta\ \mathrm{and}\ h$, where $n_{l_L}=n_{\tau_L}+n_{\nu_{\tau,L}}$ and $n_\delta=n_{c_R}-n_{b_R}$. The diffusion constants $D_i$ are given by $D_{l_L}=90/T$, $D_{\tau_R}=490/T$, $D_{t_L}=D_{t_R}=D_{b_L}=7.4/T$ and $D_{h}=20/T$ \cite{Cline:2000nw,Joyce:1994zn}, and the net collision rates are \cite{deVries:2018tgs}
\bea\label{eq:netRates}
    {\cal C}_{\tau_R}^{\sss V\!I\!A} &=& \Gamma_{\sss Y}^{\sss(\tau)}\Big(\frac{n_{\tau_R}}{k_\tau}-\frac{n_{l_L}}{2k_\tau}+\frac{n_h}{k_h}\Big) + \Gamma_{\sss M}^{\sss(\tau)}\Big(\frac{n_{\tau_R}}{k_\tau}-\frac{n_{l_L}}{2k_\tau}\Big) \nn \\
    {\cal C}_{l_L}^{\sss V\!I\!A} &=& - {\cal C}_{\tau_R}^{\sss V\!I\!A} \nn \\
    {\cal C}_{t_L}^{\sss V\!I\!A} &=& \frac{1}{2}\Gamma_{\sss Y}^{\sss (q)}\Big( \frac{n_{t_L}}{k_q}-\frac{n_{t_R}}{k_q}+\frac{n_h}{k_h} \Big) + \Gamma_{\sss M}^{\sss (q)}\Big( \frac{n_{t_L}}{k_q}-\frac{n_{t_R}}{k_q} \Big) \nn\\
    && + \Gamma_{\sss W}\Big( \frac{n_{t_L}}{k_q}-\frac{n_{b_L}}{k_q} \Big) + {\tilde \Gamma}_{\sss S\!S}[n_i] \nn\\
    {\cal C}_{b_L}^{\sss V\!I\!A} &=& \frac{1}{2}\Gamma_{\sss Y}^{\sss (q)}\Big( \frac{n_{b_L}}{k_q}-\frac{n_{t_R}}{k_q}+\frac{n_h}{k_h} \Big) \nn\\
    && - \Gamma_{\sss W}\Big( \frac{n_{t_L}}{k_q}-\frac{n_{b_L}}{k_q} \Big) + {\tilde \Gamma}_{\sss S\!S}[n_i]\nn\\
    {\cal C}_{t_R}^{\sss V\!I\!A} &=& -{\cal C}_{t_L}^{\sss V\!I\!A} - {\cal C}_{b_L}^{\sss V\!I\!A} + {\tilde \Gamma}_{\sss S\!S}[n_i] \nn\\
    {\cal C}_{\delta}^{\sss V\!I\!A} &=& 0 \nn \\
    {\cal C}_{h}^{\sss V\!I\!A} &=& \Gamma_{\sss Y}^{\sss (q)}\Big( \frac{n_{t_L}+n_{b_L}}{2k_q}-\frac{n_{t_R}}{k_q}+\frac{n_h}{k_h} \Big) + \Gamma_h\frac{n_h}{k_h} \nn\\
    && -\Gamma_{\sss Y}^{\sss(\tau)}\Big(\frac{n_{l_L}}{2k_\tau}-\frac{n_{\tau_R}}{k_\tau}-\frac{n_h}{k_h}\Big)
\eea
where we omitted the superscript $V\!I\!A$ on all the collision rates to simplify the notation. Due to the small $\tau$ Yukawa rate (and weak sphaleron rate, which is assumed to be negligible here), one can see that the lepton subsystem is approximately decoupled from the quarks + Higgs subsystem. One can then simplify the network to the species  directly activated by a source term and neglect the others.

Assuming that the unsourced particles of the first two families and $b_R$ behave similarly and satisfy $n_L^{(1,2)}=-n_R^{(1,2)}$, and imposing conservation of $B$ and $L$ separately, one finds that
\bea
n_{q_L}^{(1,2)}&=&n_{t_L}+n_{t_R}+n_{b_L}+n_{\delta}, \nn\\
n_l^{(1,2)}&=&0.
\eea
One can then write the explicit form of the weak and strong sphaleron collision terms as
\bea
    {\tilde\Gamma}_{\sss W\!S}[n_i] &=& \frac{\Gamma_{\sss W\!S}}{2}\sum_i^3\left[\frac{3}{k_q}\left(n_{q_{L,1}}^{\sss(i)}+n_{q_{L,2}}^{\sss (i)}\right)+\frac{2}{k_\tau}n_{l_L}^{\sss(i)} \right] \nn \\
    &=& \frac{\Gamma_{\sss W\!S}}{2}\left( 5n_{t_L}+5n_{b_L}+4n_{t_R}+4n_{\delta}+n_{l_L} \right) \nn\\
    {\tilde\Gamma}_{\sss S\!S}[n_i] &=& \frac{\Gamma_{\sss S\!S}}{k_q}\sum_i^3\left(n_{q_{L,1}}^{\sss(i)}+n_{q_{L,2}}^{\sss (i)}-n_{q_{R,1}}^{\sss(i)}-n_{q_{R,2}}^{\sss(i)} \right) \nn \\
    &=& \frac{\Gamma_{\sss S\!S}}{k_q}\left( 10n_{t_L}+10n_{b_L}+8n_{t_R}+8n_{\delta}\right)
\eea

As previously discussed, the WKB formalism yields a set of first-order equations describing the chemical potential and velocity perturbation of each species. It takes the general form
\be
A_i\left({\mu_i\atop u_i}\right)' +(m^2_i)' B_i\left({\mu_i\atop u_i}\right)
    -{\cal C}_i^{\sss W\!K\!B} = \left(v_w S_{{\sss W\!K\!B},i}\atop S_{{\sss W\!K\!B},i}\right),
\ee
where $A_i$ and $B_i$ are $2\times2$ model-independent matrices given in Ref.\ \cite{Cline:2020jre}, and again, $\mu_{l_L}=\mu_{\tau_L}+\mu_{\nu_{\tau,L}}$. Since the charm quark is never sourced in the WKB method, there is no need for the $\delta$ species so we can set $\mu_\delta=u_\delta=0$. The collision terms are $\mathcal{C}_i^{\sss W\!K\!B}=\left(C_i^{(1)},C_i^{(2)}\right)^\intercal$, where the $C_i^{(1)}$ are equal to the net collision rates given in Eq.\ (\ref{eq:netRates}) multiplied by $6/k_i$, and $C_i^{(2)}=-u_i/(3D_i)-v_w C_i^{(1)}$. The coefficients $C_i^{(1)}$ can be written in terms of the WKB collision rates (see Eq.\ (\ref{eq:WKBrates})) and the chemical potentials as
\bea
    C_{\tau_R}^{(1)} &=& \Gamma_{\sss Y}^{\sss(\tau)}(2\mu_{\tau_R}-\mu_{l_L}+2\mu_h) + 2\Gamma_{\sss M}^{\sss(\tau)}(\mu_{\tau_R}-\mu_{l_L}) \nn \\
    C_{l_L}^{(1)} &=& - C_{\tau_R}^{(1)} \nn \\
    C_{t_L}^{(1)} &=& \Gamma_{\sss Y}^{\sss (q)}( \mu_{t_L}-\mu_{t_R}+\mu_h ) + 2\Gamma_{\sss M}^{\sss (q)}( \mu_{t_L}-\mu_{t_R}) \nn\\
    && + \Gamma_{\sss W}( \mu_{t_L}-\mu_{b_L} ) + {\bar \Gamma}_{\sss S\!S}[\mu_i] \nn\\
    C_{b_L}^{(1)} &=& \Gamma_{\sss Y}^{\sss (q)}( \mu_{b_L}-\mu_{t_R}+\mu_h ) \nn\\
    && - \Gamma_{\sss W}( \mu_{t_L}-\mu_{b_L}) + {\bar \Gamma}_{\sss S\!S}[\mu_i] \nn\\
    C_{t_R}^{(1)} &=& -C_{t_L}^{(1)} - C_{b_L}^{(1)} + {\bar \Gamma}_{\sss S\!S}[\mu_i] \nn\\
    C_h^{(1)} &=& \frac{3}{2}\Gamma_{\sss Y}^{\sss (q)}( \mu_{t_L}+\mu_{b_L}-2\mu_{t_R}+2\mu_h) + \Gamma_h\mu_h \nn\\
    && -\frac{1}{2}\Gamma_{\sss Y}^{\sss(\tau)}(\mu_{l_L}-2\mu_{\tau_R}-2\mu_h)
\eea
with 
\bea
    {\bar\Gamma}_{\sss W\!S}[\mu_i] &=& \frac{\Gamma_{\sss W\!S}}{2}\left( 15\mu_{t_L}+15\mu_{b_L}+12\mu_{t_R}+\mu_{l_L} \right) \nn\\
    {\bar\Gamma}_{\sss S\!S}[\mu_i] &=& \Gamma_{\sss S\!S}\left( 10\mu_{t_L}+10\mu_{b_L}+8\mu_{t_R}\right)\hspace{1cm}
\eea
We omitted the superscript $W\!K\!B$ on all the collision rates to simplify the notation.

\end{appendix}

\bibliography{ref}
\bibliographystyle{utphys}

\end{document}